# Optimal STEM Convergence Angle Selection using a Convolutional Neural Network and the Strehl Ratio

Noah Schnitzer[1,2], Suk Hyun Sung[1], Robert Hovden[1,3]

[1.]Department of Materials Science and Engineering, University of Michigan, Ann Arbor, MI 48019, USA

[2.]Department of Materials Science and Engineering, Cornell University, Ithaca, NY 14853, USA

[3.]Applied Physics Program, University of Michigan, Ann Arbor, MI 48109, USA

**Abstract**

Selection of the correct convergence angle is essential for achieving the highest resolution imaging in scanning transmission electron microscopy (STEM). Use of poor heuristics, such as Rayleigh's quarter-phase rule, to assess probe quality and uncertainties in measurement of the aberration function result in incorrect selection of convergence angles and lower resolution. Here, we show that the Strehl ratio provides an accurate and efficient to calculate criteria for evaluating probe size for STEM. A convolutional neural network trained on the Strehl ratio is shown to outperform experienced microscopists at selecting a convergence angle from a single electron Ronchigram using simulated datasets. Generating tens of thousands of simulated Ronchigram examples, the network is trained to select convergence angles yielding probes on average 85% nearer to optimal size at millisecond speeds (0.02% human assessment time). Qualitative assessment on experimental Ronchigrams with intentionally introduced aberrations suggests that trends in the optimal convergence angle size are well modeled but high accuracy requires extensive training datasets. This near immediate assessment of Ronchigrams using the Strehl ratio and machine learning highlights a viable path toward rapid, automated alignment of aberration-corrected electron microscopes.

**Introduction**

The modern breakthrough in scanning transmission electron microscope (STEM) resolution can be attributed to both the multi-pole electromagnetic lenses that correct aberrations and complex software that assess the aberrations to be corrected. However, the final imaging resolution of a well aligned microscope critically depends on the convergence angle set by the size of the objective aperture. Determining optimal convergence angles remains difficult for experts even with computer aid. Firstly, the experimental aberration function is unknown. Secondly, even with perfect knowledge of the aberration function there is not a straightforward and efficient method of identifying the optimum convergence angle and aperture placement that maximizes resolution. A number of approaches are currently used to estimate the aberration function and select a convergence angle, but they are limited by the tolerances of their estimates. Even worse they use heuristics such as Rayleigh's quarter-phase rule that fail to accurately predict the best convergence angle even if the aberration function is exactly known.

Here, we present a deep convolutional neural network (CNN) for predicting the optimal convergence angle for STEM imaging with the Strehl ratio. CNNs have been shown to have remarkable performance at image analysis tasks (LeCun et al., 2015) such as classification (Krizhevsky et al., 2012), encoding and decoding (Badrinarayanan et al., 2017), and regression (Lathuiliére et al., 2019; Mahendran et al., 2017), including recent applications in electron microscopy (Ede & Beanland, 2019; Xu & LeBeau, 2018; Zhang et al., 2020). The Strehl ratio is an accurate and efficient to calculate metric for probe quality that straightforwardly incorporates into an objective function for optimization of the STEM probe. The Strehl ratio quantifies the reduction in peak probe intensity caused by all aberrations. Using a CNN to assess the Strehl ratio from the Ronchigram, we report strong performance at convergence angle selection exceeding that of trained microscopists on a dataset of simulated Ronchigrams. This machine learning approach is qualitatively validated on electron Ronchigrams collected experimentally on an aberration-corrected STEM. This approach not only improves convergence angle selection, but provides the high-speed and accurate assessment required for rapid automated STEM alignment.

**Background**

The electron Ronchigram is a vital tool for assessing and correcting aberrations for STEM. The Ronchigram takes its name from a test devised for assessing aberrations in light optical systems, the "Ronchi Test." Developed by Italian physicist Vasco Ronchi in 1923, it uses a linear grating spaced ~100x the wavelength of light to form an interference pattern which encodes the aberrations of the optical system (Ronchi, 1923, 1964). In the 1970s, Cowley and other pioneers in STEM recognized that crystalline and amorphous atomic structures form gratings in the electron microscope, and an analogous interference pattern encoding aberrations could be formed with STEM's convergent beam to aid in alignment (Cowley, 1979a, 1979b, 1980; Lin & Cowley, 1986).

Assessing the aberrations in STEM is vital for improving imaging resolution and contrast (Kirkland, 2011). Lens aberrations cause a shift in the phase of the electron wavefunction across the objective aperture. Aberrations are deviations from a perfect spherical wave, and an aberration function describing the phase shift can be compactly expanded in terms of the radial ($\alpha$) and azimuthal ($\phi$) angles using the Krivanek notation (Krivanek et al., 1999): $\chi(\alpha, \phi) = \frac{2\pi}{\lambda} \sum_{n,m} \frac{C_{n,m} \alpha^{n+1} \cos(m(\phi - \phi_{n,m}))}{n+1}$ where $C_{n,m}$ and $\phi_{n,m}$ describe a geometric aberration's magnitude and orientation, $m$ is the order of rotational symmetry (0 for cylindrically symmetric aberrations, otherwise $\frac{2\pi}{m}$ is the smallest angle such that the phase shift of the aberration is equivalent), $n$ is the order of the aberration, and $\lambda$ is the wavelength of the electron beam. This phase shift results in larger probe tails and a smaller maximum intensity for the STEM probe (Kirkland, 2011).

In an uncorrected STEM, the achievable resolution is limited by third order spherical aberration ($C_{3,0}$), which grows with the radial angle. To achieve the best possible resolution, the spherical aberration is balanced with defocus ($C_{1,0}$) and an aperture placed at the objective lens blocks highly-aberrated portions of the beam (Scherzer, 1949; Krivanek et al. 2008). Near the optical axis, a minimally aberrated portion of the electron beam passes through the aperture. The size of the objective aperture directly relates to the STEM convergence angle. Using a smaller aperture reduces the phase shift of the wavefunction, but too small of an aperture limits resolution due to diffraction. The aperture sets a diffraction limited resolution of $d = 0.61\lambda/\alpha$ for a convergence semi-angle $\alpha$ per Lord Rayleigh's resolution criterion (Rayleigh, Lord, 1896). It is essential that for the best imaging conditions the convergence angle (i.e. aperture size) be chosen such that it balances the limits imposed by aberrations and diffraction (Crewe, 1982; Weyland & Muller, 2005).

Currently, Rayleigh's quarter-wave rule is widely used in electron microscopy—an assumption that less than a quarter wavelength phase shift due to spherical aberration will permit good imaging quality (Rayleigh, Lord, 1879). Quantitatively, Rayleigh's quarter-wave rule corresponds to a 20% decrease in the peak intensity of the probe (Mahajan, 1982). Using the uncorrected instrument's known spherical aberration, it is straightforward to calculate the largest convergence angle with less than a quarter wavelength phase shift as derived by Weyland and Muller (Weyland & Muller, 2005). However, with the popularization of aberration correction, spherical aberration no longer dominates (Krivanek et al., 1997). Instead, higher-order and parasitic geometric aberrations—those arising from correction with non-spherically symmetric multipole elements—as well as chromatic aberration limit the resolution with complex functional form (Haider et al., 2000). These aberrations can still be mitigated using a smaller convergence angle; however, the same tradeoff with the diffraction limit remains, and selecting the correct convergence angle to optimize resolution becomes more difficult.

Firstly, the aberration function is unknown. A variety of techniques exist to attempt to estimate the aberration function using various signals available in STEM. Some of the most effective techniques leverage the electron Ronchigram. Most follow a similar workflow: segmenting the Ronchigram (or often multiple Ronchigrams with a parameter intentionally manipulated) into regions of approximately constant behavior, applying an image processing technique to each segment, and fitting extracted values to the aberration function. For instance, the technique used on JEOL microscopes involves applying an autocorrelation to each segment of two Ronchigrams separated by a known defocus (Sasaki et al., 2010; Sawada et al., 2008), while on Nion microscopes Ronchigrams are captured with various shifts, and the segments (roughly corresponding to beam angle) are cross-correlated (Dellby et al., 2001). The aberration function can also be measured by taking the Fourier Transform of segments of the Ronchigram (Lupini et al., 2010). Assessment of the aberration function in real space is also possible (Wong et al., 1992), for instance microscopes with CEOS correctors use the ADF-STEM signal (Lazic et al., 2014; Janssen et al., 2015; Henstra & Tiemeijer, 2018). However, this real-space approach requires inserting a known calibration specimen and relatively slow image acquisition. All of these techniques only provide estimates of the aberration function, and the iterative alignment process is slow to converge. In practice, human assisted assessment is required to reliably align the beam to within the tolerances required for high-convergence angles. Ultimately, the microscope user will decide when the microscope is aligned well enough for their target resolution and convergence angle. However, these procedures can leave residual aberrations which can have a significant impact on imaging resolution or create insidious probe tails if unaccounted for in convergence angle selection (Kirkland, 2011).

**Assessing Probe Quality with the Strehl Ratio**

Secondly, even with perfect knowledge of the aberration function there is not a straightforward and efficient method of finding the optimum convergence angle to achieve the best imaging resolution. The two most prominent methods are inspired by Rayleigh's quarter-wave rule, and involve either selecting the largest convergence angle such that each individual aberration has less than a quarter wavelength phase shift (Figure 1a, orange lines) (Dellby et al., 2001; Kirkland, 2018; Sasaki et al., 2010) or such that the total aberration function (Figure 1a, black line) (Müller et al., 2006) varies by less than a quarter wave in each direction (Figure 1a, yellow box). These heuristics are intuitively related to the aberration function phase shift and fast to calculate.

Unfortunately, applying a quarter-wave rule does not result in the sharpest probe as the quarter-wave rule is only optimal for rotationally symmetric, spherical aberrations. On aberration-corrected microscopes, parasitic higher-order aberrations that are not rotationally symmetric greatly distort the beam. A quarter wave shift from non-spherical aberrations (e.g. astigmatism, coma) will no longer reduce the peak probe intensity by 20%, meaning that the quarter-wave rule will not quantitatively relate the probe quality (Mahajan, 1982). In STEM the problem is complicated in that a number of significant aberrations of various orders and symmetries are present which not only individually cannot be assessed with the quarter-wave rule, but will also interfere with one another constructively or destructively based on their symmetry and orientation. This implies that existing techniques in use are not selecting the optimal convergence angle, even for a perfectly measured aberration function.

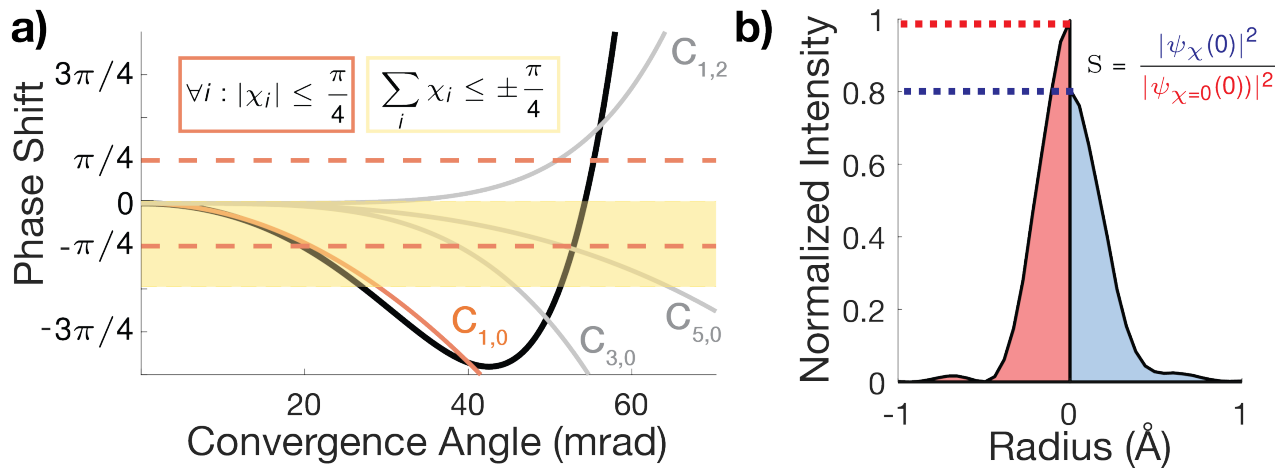

**Figure 1:** Criteria for selecting the optimal convergence angle for STEM. **a)** Phase shift is plotted against convergence angle for a complex aberration function, the total aberration function phase shift (yellow), and individual aberration phase shift (orange) criteria are indicated with boxed regions. **b)** Probe radial intensity is plotted for a non-aberrated (left) and aberrated (right) system with a Strehl ratio of 0.8, and a convergence angle of 23 mrad.

An alternative is to select the optimum convergence angle from the real space probe rather than the aberration function. Rose proposed selecting the convergence angle such that it minimizes the diameter which encloses 59% of the probe current (Rose, 1981). In the diffraction limited case this corresponds to the first zero of the first order Bessel function ($J_1$) and bears a direct relationship

Schnitzer 7greatly distort the beam. A quarter wave shift from non-spherical aberrations (e.g. astigmatism, coma) will no longer reduce the peak probe intensity by 20%, meaning that the quarter-wave rule will not quantitatively relate the probe quality (Mahajan, 1982). In STEM the problem is complicated in that a number of significant aberrations of various orders and symmetries are present which not only individually cannot be assessed with the quarter-wave rule, but will also interfere with one another constructively or destructively based on their symmetry and orientation. This implies that existing techniques in use are not selecting the optimal convergence angle, even for a perfectly measured aberration function.

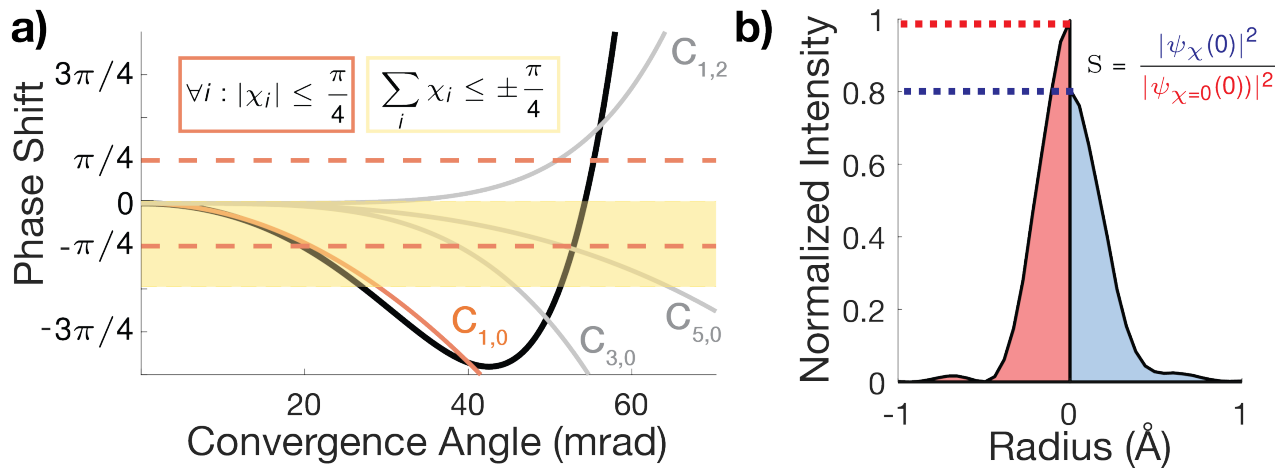

**Figure 1:** Criteria for selecting the optimal convergence angle for STEM. **a)** Phase shift is plotted against convergence angle for a complex aberration function, the total aberration function phase shift (yellow), and individual aberration phase shift (orange) criteria are indicated with boxed regions. **b)** Probe radial intensity is plotted for a non-aberrated (left) and aberrated (right) system with a Strehl ratio of 0.8, and a convergence angle of 23 mrad.

An alternative is to select the optimum convergence angle from the real space probe rather than the aberration function. Rose proposed selecting the convergence angle such that it minimizes the diameter which encloses 59% of the probe current (Rose, 1981). In the diffraction limited case this corresponds to the first zero of the first order Bessel function ($J_1$) and bears a direct relationship to image contrast and resolution (Haider et al., 2000). However, it is expensive to compute—finding the minimum convergence angle requires calculation of many probes each of which must be integrated to find the 59% diameter. While this is prohibitive for use in real time convergence angle selection or generation of very large datasets, it is an accurate baseline for assessing the quality of other less expensive heuristics. In Figure 2a,b, using Kirkland's convention the diameter which includes 50% of the probe current is termed the "probe size," and is used to evaluate the other heuristics due to its bearing on imaging resolution and contrast (Kirkland, 2011). A comparison of the 50% and 59% probe current diameters as a function of convergence angle is shown in Supplemental Figure 1.

The Strehl ratio is a probe quality metric that outperforms the quarter-wave rule heuristics in assessing the optimal convergence angle. The Strehl ratio is the ratio of peak intensity between an aberrated (Figure 1b, blue) and non-aberrated (Figure 1b, red) probe, which is frequently used in the adaptive optics community as a robust measure of probe quality that can be easily approximated and interpreted (Burke et al., 2015; Porro et al., 1999). Notably, a Strehl ratio of 0.8 corresponds to a 20% decrease in peak probe intensity, providing a quantitative analogue to the Rayleigh quarter-wave rule that can be applied to non-spherically symmetric, complex probes (Mahajan, 1982). Here, the Strehl ratio is used to define a criterion for convergence angle selection: selecting the largest convergence angle such that the Strehl ratio is at least 0.8 (Figure 1b). As shown in Figure 2a, for a randomly generated aberration function this criterion (blue line) comes closer to the minimum probe size than the phase shift heuristics (yellow, orange lines). While the absolute difference in probe size is small (picometers), it demonstrates a systematic loss of probe quality. The convergence angles selected by the 0.8 Strehl ratio, the total aberration phase shift, and the individual aberration phase shift are compared in Figure 2c and Supplemental Figure 2.

Over hundreds of randomly generated aberration functions tuned to correspond to those in an aberration-corrected STEM and scaled to span a broad range of optimum convergence angles, the 0.8 Strehl ratio criterion closely approximates the minimum probe size resulting in probes on average less than a picometer larger (Figure 2b). The Strehl ratio for a probe with a given convergence angle is calculated without iteration, allowing the 0.8 Strehl ratio convergence angle to be found in a fraction of the time taken to find the minimum probe size convergence angle. Its quick calculation and high accuracy make the Strehl ratio an ideal single-value heuristic for convergence angle selection. However, like all of the heuristics discussed, the Strehl ratio is calculated from the estimated aberration function. By training a CNN to predict the Strehl ratio convergence angle directly from the Ronchigram, this limitation can be bypassed.

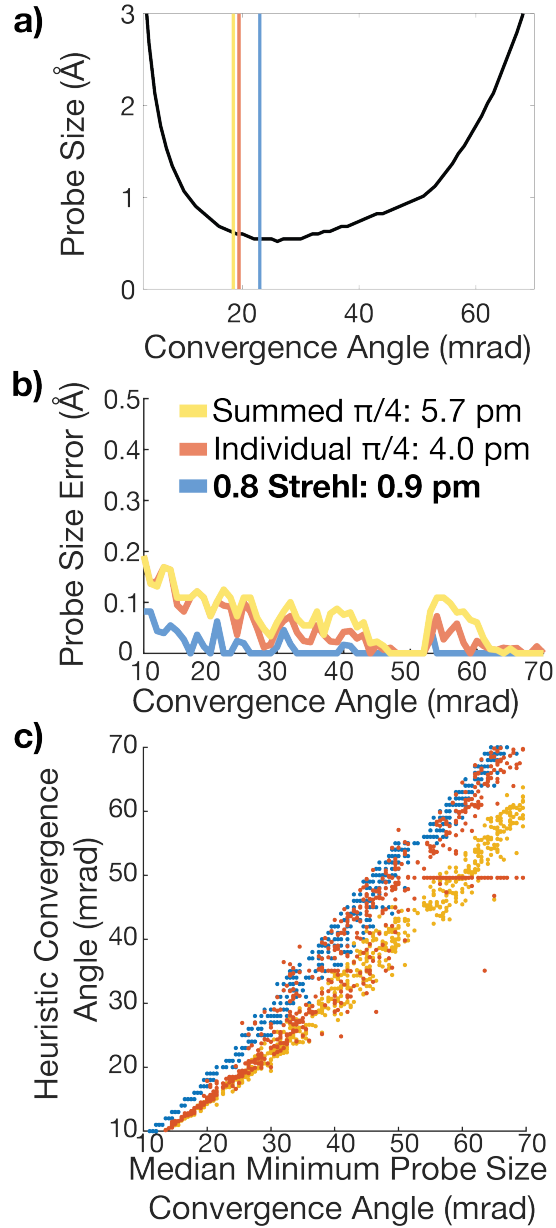

**Figure 2:** Comparison of convergence angle selection criteria. **a)** Probe size (50% probe current diameter) is plotted against convergence angle for a complex non-spherically symmetric aberration function, with probe sizes selected by 0.8 Strehl ratio (blue), total aberration function phase shift (yellow), and individual aberration phase shift (orange) criteria. **b)** Plot of average error in probe size versus the minimum for each criteria, over a set of 664 aberrations from 10 to 70 mrad. The values in the legend are the mean probe size error of the 664 aberrations. Compare to Figure 4c for CNN and human results on same set. **c)** Convergence angles selected by the Strehl and phase

shift criteria plotted against the median convergence angle which results in the minimum 50% probe current diameter. The line of individual aberration phase shift selections at 49.6 mrad correspond to a π/4 phase shift in defocus, which is accentuated in this dataset as defocus was set to compensate for the other aberrations in discrete 2 Å steps.

**Materials and Methods**

**Convolutional Neural Network for Ronchigrams**

In this work, we develop a CNN for selection of the optimal convergence angle from the electron Ronchigram. Training our network on a large dataset of simulated Ronchigrams labeled with the 0.8 Strehl ratio convergence angle, we are able to develop a model which approximates the unknown non-linear function mapping from the Ronchigram to the convergence angle.

Deep neural networks have demonstrated state of the art performance on a wide variety of tasks in recent years. In supervised learning, particular success has been seen for problems where training on very large, low noise data sets is possible. For image analysis problems, CNNs have received the greatest interest and the most promising results. As detailed by LeCun et al., using shared multidimensional convolutional filters CNNs retain spatial relationships and have a reduced number of learnable parameters in comparison to traditional neural networks (LeCun et al., 1998). Hence, the CNN is a performative, highly tunable, and relatively easy to train architecture ideal for application to image analysis problems in which large labeled datasets can be provided.

We based our architecture on AlexNet (Krizhevsky et al., 2012), borrowing features such as ReLU activation and max pooling layers. Unlike AlexNet, our model is for a single variable regression, rather than multi-class classification — it predicts a single continuous value rather than probabilities for a large number of discrete classes — so smaller fully connected layers were used, and no activation was used for the final layer as its output is directly taken as the normalized optimal convergence angle. As illustrated in Figure 3a, the input layer was sized 512x512x1, corresponding to a typical binned CCD camera image, and all convolutional layers were padded to keep their output the same size as their input. Additionally, hyperparameters such as filter dimensions and number of layers were optimized to enhance convergence to a minimum of our objective function, the root mean squared error of the predicted convergence angle to the 0.8 Strehl ratio convergence angle. Our final network had approximately the same number of parameters as AlexNet, and slightly improved performance in comparison to transfer learning on a pretrained AlexNet with input images mapped to RGB with equal intensity for each channel and resized to AlexNet's 227x227 pixel dimensions. Training and validation performance of the final network are shown along with filter morphology in Figure 3b. MATLAB's Deep Learning Toolbox was used to implement the network and compare transfer learning with AlexNet.

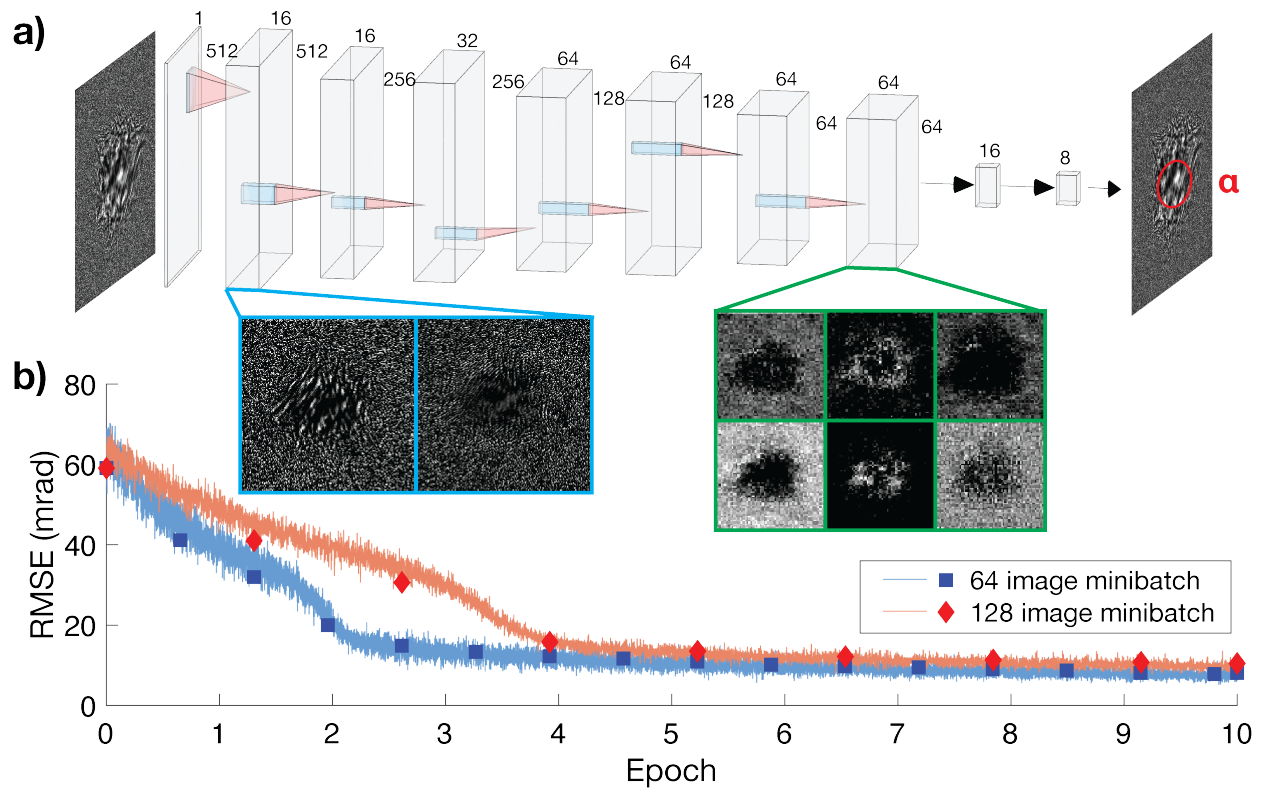

**Figure 3:** Convolutional Neural Network Architecture. **a)** Schematic of network, illustrating convolutional layers and filter sizes along with selected activation maps of the first and last convolutional layers. **b)** Network error in convergence angle estimation across 10 epochs of training for two different minibatch sizes (solid lines), as well as performance on a validation set (square, diamond markers). The minibatch size affects convergence rate but does not significantly affect the converged error rate.

**Training and Testing on Ronchigram Simulation**

Our network was trained on simulated images of electron Ronchigrams labeled with the 0.8 Strehl ratio convergence angle. Training on simulated data was made necessary by the number of training examples required for the network to converge, and the difficulty of generating trustworthy labels. Here, 98,000 simulated images were provided; however, image classification datasets can frequently contain millions of labeled images, such as ImageNet (Deng et al., 2009). Collecting such large numbers of experimental Ronchigrams would require a prohibitive amount of time on an aberration-corrected STEM, and errors in aberration function estimation would make trustworthy labeling impossible.

Simulation of the electron Ronchigram can easily be achieved with high accuracy by most multislice simulation packages (Kirkland, 2010; Barthel, 2018; Ophus, 2017). However, these simulations are not typically optimized for this calculation: for the ultra-thin amorphous specimens typically used for tuning many effects such as multiple scattering need not be accounted for. With efficient approximations for the electron optics, a large part of the remaining computational cost of the calculation comes from generating the amorphous specimen potential. It has previously been shown that substitution of an oversampled white noise grating can be used to roughly approximate the specimen potential of a thin amorphous material with significantly reduced computational cost

(Schnitzer et al., 2019).

Ronchigrams were simulated with only geometric aberrations, resulting in a probe wave function $\psi_p(\mathbf{k}) = e^{-i\chi(\mathbf{k})}$. Under the eikonal approximation the transmission function can be described as $\psi_t(\mathbf{x}) = FT\{e^{-i\chi(\mathbf{k})}\} \cdot e^{-i\sigma V(\mathbf{x})}$ for an interaction parameter $\sigma$ and a specimen potential function or noise grating $V(\mathbf{x})$. The Ronchigram is the square modulus of the transmission function in the diffraction plane $g(\mathbf{k}) = |FT\{\psi_t(\mathbf{x})\}|^2$. The Strehl ratio (S) is easily calculated by comparing the peak intensities of an aberrated and non-aberrated probe with the same convergence angle, $S = \frac{\psi_\chi(0)}{\psi_{\chi=0}(0)}$, or equivalently by comparing the peak intensity of an aberrated probe with its total intensity (Mahajan, 1982). A monochromatic point source approximation was used. This works best at higher beam voltages where a thin specimen approximation is valid and contributions to chromatic aberrations are small. A finite source size, as well as chromatic aberration, add a damping envelope that attenuates interference fringes in the Ronchigram (Lupini et al. 2010; Dwyer et al. 2010).

Aberration functions for the training data set were randomly generated. The magnitude and angle for each term of the aberration functions were generated from a uniform distribution scaled to correspond to the relative magnitudes of aberrations in aberration-corrected STEM (Supp. Table 1). The aberration functions were further scaled to approximately uniformly span a range of optimum 0.8 Strehl convergence angles of 1 to 105 mrad. All Ronchigrams were simulated with 1024x1024 pixel wavefunctions, out to a maximum radial angle of 180 mrad with a 128 mrad convergence angle in double precision. After simulation, Ronchigrams were slightly post-processed before being input into the network: images were cropped to a square with a field-of-view of 180 mrad (512x512 pixels), normalized to the maximum pixel value, and stored in 8-bit grayscale to minimize storage use. This post-processing pipeline leverages no knowledge of the simulation parameters except the pixel dimension of the convergence angle and was easily replicated for experimental Ronchigrams.

The network was trained with stochastic gradient descent with an initial learning rate of $1 \times 10^{-5}$ which was reduced to $1 \times 10^{-6}$ after 50 epochs and $1 \times 10^{-7}$ after 100 epochs, and a momentum of 0.9. Training was ended after 101 epochs, at which point both training and validation loss had plateaued. $L_2$ regularization was used with a regularization parameter of $1 \times 10^{-4}$. Mini-batches of 64 Ronchigrams were used to speed convergence in training (Figure 3b). Hyperparameters were manually tuned to optimize speed of training convergence and the final value of the objective function on the validation set– the root mean squared error of the predicted convergence angle to the 0.8 Strehl ratio aperture size. 98,000 simulated Ronchigrams with randomly generated aberration functions were used to train the network, 1,000 were used to validate the network during training to inform hyperparameter selection, and 1,000 were used to test the trained network, all with identical simulation parameters and similar distributions of 0.8 Strehl ratio convergence angles. Of the 1,000 Ronchigrams in the test set, Figures 2b and 4c include 664 with minimum probe size convergence angles between 10 and 70 mrad. In addition to being tested on by the final network architecture, the test set was assessed by the authors of this manuscript who estimated optimal convergence angles by drawing circles over the simulated Ronchigrams.

**Experimental Ronchigram Collection**

Electron Ronchigrams were acquired experimentally on a double corrected JEOL 3100R05 with a Gatan Ultrascan 1000 CCD TV camera. A typical alignment operating condition was used:

300 keV with a 150 μm objective aperture (110 mrad convergence angle), 8 cm camera length, and one second acquisition time. Ronchigrams were formed with 50 nm thick amorphous Silicon Nitride TEM grids (Norcada Inc.). Defocus and two-fold astigmatism were intentionally introduced using the JEOL COSMO software, and series of Ronchigrams with varying defocus were acquired with custom software. Acquired Ronchigrams were post-processed with practically the same procedure as the simulated Ronchigrams: images were cropped to the largest square inscribed by the objective aperture, down-sampled to 512x512 pixels, normalized to the maximum pixel value, and converted to 8-bit grayscale.

**Results**

**Performance on Simulated Ronchigrams**

The performance of the CNN was assessed through comparison on a set of 1000 simulated Ronchigrams with the 0.8 Strehl ratio convergence angle, the minimum probe size convergence angle, and the choice of experienced microscopists. As the aberration functions for these simulated Ronchigrams were exactly known this comparison could be precisely quantified.

The CNN strongly approximated the 0.8 Strehl ratio criteria for convergence angle selection. The root mean squared error of the predicted convergence angle was less than 4 mrad on both the training and test sets (Suppl. Table 2), approaching but falling short of the 0.35 mrad per pixel sampling rate of the Ronchigrams. Thus, the CNN's performance is not limited by the sampling of its input, but by either the structure and training of the CNN or limitations on the precision to which the Strehl ratio convergence angle can be predicted by a single electron Ronchigram.

The CNN's predictions were also compared with the minimum possible probe sizes, as assessed by the 50% probe current heuristic. As shown in Figure 4a and 4b, while the convergence angle predicted by the CNN did not always give have the smallest possible probe size, the difference in size was typically small (< 10 pm). Comparing the mean absolute error in probe size of the 0.8 Strehl ratio criteria (Figure 2b, blue line) and the CNN's selections (Figure 4c, purple line), the absolute errors across the full set of Ronchigrams are similar, and neither shows any trend with convergence angle. This indicates that the CNN approximates the minimum probe size from the Ronchigram nearly as well as the 0.8 Strehl ratio does provided with the complete aberration function.

Most dramatically, the convergence angles predicted by the CNN resulted in significantly smaller probes than those selected by experienced microscopists: across the set of 1000 Ronchigrams, the network's selections were on average 85% closer to the minimum probe size, and took just 0.02% of the time to find (Figure 4c). Considering the probe size as a function of convergence angle, it is clear that the network is significantly better than the human eye at analyzing the Ronchigram to find the minimum point balancing the diffraction and aberration limits—it predicts a closer convergence angle to the minimum, resulting in a measurably smaller probe size (Figure 4a,b).

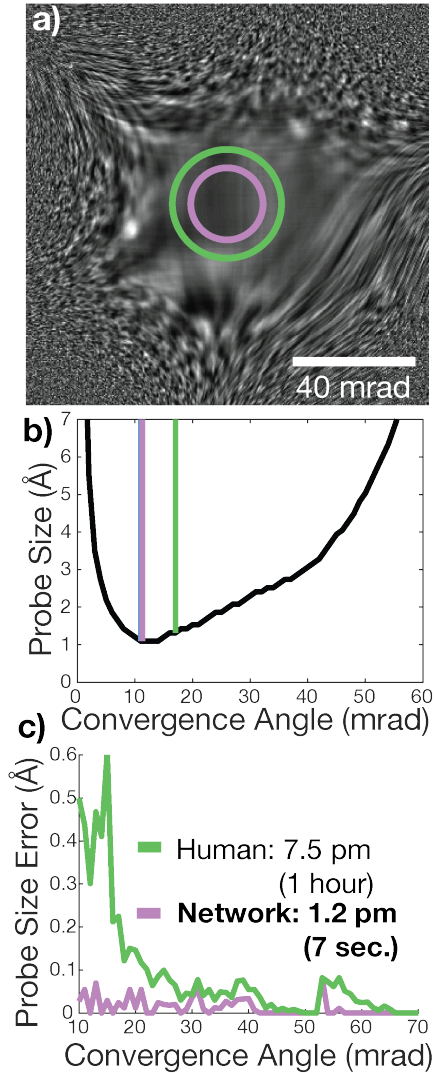

**Figure 4:** Network and microscopist performance on the held out simulated test set. **a)** Example simulated Ronchigram, with convergence angles selected by a microscopist (green) and network (purple). **b)** Plot of probe size as a function of convergence angle for the Ronchigram in (a). The green line indicates the microscopist's guess, and the purple line the network's which lies closer to the minimum of the plot. **c)** Trend in error in probe size compared to the minimum possible probe size (50% probe current diameter) for the network and microscopists with the optimum convergence angle. The values in legend are the mean probe size error and total assessment time for 664 Ronchigrams simulated with aberrations with minimum probe sizes between 10 and 70 mrad. Compare to Figure 2b for 0.8 Strehl ratio and phase shift heuristics results on same set of aberration functions.

**Performance on Experimental Ronchigrams**

In order to assess the CNNs capability for online use on a STEM, the network was also assessed against Ronchigrams acquired experimentally on an aberration-corrected instrument. Assessment of performance on experimental Ronchigrams is more difficult as the aberration function is unknown, meaning no ground truth is available.

However, a qualitative assessment is possible by considering the trends in the network's predictions with intentionally applied low-order aberrations. Figure 5a plots the CNN's convergence angle prediction as a function of applied defocus, optimal defocus by manual observation was estimated to be close to 0 nm. The increase in CNN predicted convergence angle close to this value follows the expected trend for the 0.8 Strehl ratio criteria. Likewise, intentionally applying two-fold astigmatism, the network's predicted convergence angle was smaller (Figure 5c) than the original state (Figure 5b). However, under manual inspection the convergence angle predicted for the intentionally stigmated Ronchigram and those at large defocuses were too large, suggesting that the network predictions reached an incorrect floor on the experimental data. Given that the network was trained entirely on synthetic data and no preprocessing was supplied except trivial input resizing and normalization, this limited performance is unsurprising.

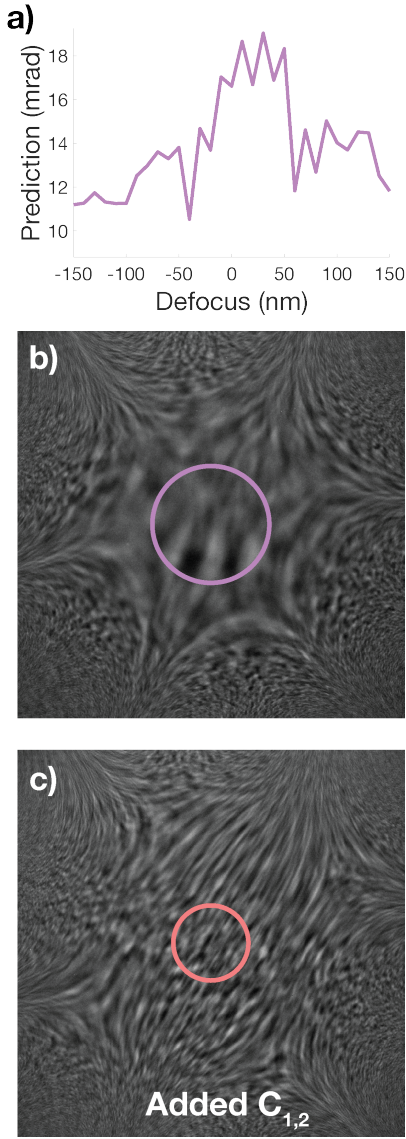

**Figure 5:** Network results on experimental electron Ronchigrams. **a)** Trend in CNN prediction with applied defocus. **b)** Experimental Ronchigram with CNN predicted convergence angle superimposed. **c)** Experimental Ronchigram with two-fold astigmatism intentionally added, smaller CNN prediction superimposed.

## Discussion

This work presents a new approach for Ronchigram analysis that maps a single electron Ronchigram directly to a convergence angle using machine learning and the Strehl ratio to make accurate, automated assessments. This is a significant departure from common Ronchigram analysis methods which through multistep processes requiring acquisition of multiple Ronchigrams, fit the aberration function and reduce it to various parameters of interest. Furthermore, common Ronchigram analysis methods calculate the optimal convergence angle from the phase shifts due to aberrations. The quantitatively strong performance shown here on simulated Ronchigrams suggests that this direct mapping with a Strehl ratio based assessment has advantages over the current norm. A direct comparison of the network presented here to the state of the art in computational Ronchigram analysis requires presently unavailable open-source or accessible implementations of STEM correction hardware and software. Open instrument access is crucial to research and development of machine learning alignment in aberration corrected STEM.

Framing convergence angle selection as a supervised image analysis problem and using a deep convolutional neural network was motivated by the recent successes in this area of computer vision. The most substantial barrier to performing this type of work for electron microscopy lies in generating the large, low noise datasets commonly used in computer vision. Here, electron Ronchigrams were quickly simulated with arbitrary aberrations, and knowledge of the aberration function enabled accurate labeling of the images and analysis of network performance. However, even with perfect knowledge of the aberration function identifying the ideal convergence angle is non-trivial. The Strehl ratio was chosen to assess the optimum convergence angle as it gives a robust measure of probe quality consistent with accepted metrics such as the probe current diameter, but is also fast enough to compute to allow creation of large training datasets. Leveraging the Strehl ratio, a large dataset of simulated Ronchigrams was built and used to train a CNN. While some experimentation in architectures, hyperparameters, and preprocessing was performed, it was far from exhaustive. With more data, additional training, or alternative network design we might expect to see filter morphology emerge that provides better physical grounding or intuition to the neural network behavior.

Training and quantitative validation of the network is performed entirely on simulated electron Ronchigrams, as using experimental Ronchigrams would require knowledge of the unknown microscope aberration function. The CNN showed qualitatively poor performance on experimental Ronchigrams compared to its performance on simulated Ronchigrams. This is unsurprising, considering the CNN was trained entirely on simulated images that did not replicate every experimental detail. A key limitation of a CNN approach and this work is that the model's performance is closely tied to its training set. Model complexity grows quickly with additional permutations of microscope parameters such as acceleration voltages and source sizes. In this work chromatic aberration, inelastic scattering, and source size were neglected as these aggressive approximations significantly speed up Ronchigram simulations but limit the extent to which the model can be applied to real systems. A more robust approach would include multiple and inelastic scattering to handle thicker specimens and chromatic aberrations to determine convergence angles at lower beam voltages. A promising future direction for applying machine learning techniques to aberration correction will leverage unsupervised or weakly supervised learning on experimental datasets either for pre-training or as part of an end to end model. For the current approach, performance could be improved with more thorough data preprocessing and neural architecture search.

**Conclusion**

The electron microscope's ability to resolve atomic structure has been revolutionized by

Schnitzer 22

the development of aberration correction hardware which opens up the usable convergence angle. However, these correctors are constructed from a complex stack of multi-pole electromagnetic lenses that must be controlled, assessed, and adjusted using software. Substantial innovation of aberration-corrected lens design ultimately relied on the maturity of desktop computing. The next generation of high current, high-resolution, and ultra-monochromatic STEM will face an even larger parameter space that demands quick and reliable tuning of all lens components. Currently, techniques using Ronchigrams to tune these lenses are imperfect and ultimately rely on an iterative process, qualitative human adjustment, and machine assisted algorithms that acquire and analyze multiple Ronchigrams. This limits the final resolution.

The work herein suggests new heuristics for assessing beam shape that facilitate machine learning approaches to microscope alignment. Using simulated data of coherent sources on thin specimens, we show assessment is near-immediate and limited only by the acquisition time of a single Ronchigram. We present a CNN model which outperforms trained microscopists at convergence angle selection on simulated Ronchigrams—a significant result given that this task is typically left to the user or informed through poor heuristics in software. Crucially, through comparison with trustworthy probe quality metrics we find that commonly used cutoffs in the phase shift of the aberration function are insufficient to select the convergence angle. We identify the Strehl ratio as an efficient and accurate alternative. Training the CNN on large simulated datasets labeled with this strong heuristic is essential for the model's effectiveness, which we assess through comparison of the resulting probe size to the minimum possible (assessed by the diameter enclosing 50% of the probe current).

This work opens a new route toward fast, accurate, automated alignment for aberration-corrected STEM from a single Ronchigram using machine learning. We show alternative metrics

Schnitzer 23

can outperform the quarter-wave rule; both for existing and machine learning based Ronchigram assessment. Implementation can be further improved with a wider class of Ronchigram simulations that include incoherent sources and multiple scattering as well as integration of experimental data. However, convergence angle selection is just one part of the STEM alignment. To achieve fast, accurate, and automated alignment, additional sub-problems such as objective aperture placement and feedback to the corrector lenses must be improved. These goals are achievable with open scientific development of electron microscope hardware.

To aid in further inquiry in this area, the MATLAB code for Ronchigram simulation and assessment, dataset generation, network training, and network validation is available at https://github.com/noahschnitzer/ronchigram-matlab under a GPL 3.0 license.

## Acknowledgements

R. H. acknowledges support from Army Research Office, Computing Science Division (#74589CS). N.S. and S. H. S. contributed equally to this work with support from NSF grant #DMR-0723032. The authors utilized the Michigan Center for Materials Characterization (MC2) and computational resources from Advanced Research Computing at the University of Michigan, Ann Arbor as well as the GPU grant program from NVIDIA Corporation.

Xu, W. & LeBeau, J. M. (2018). A deep convolutional neural network to analyze position averaged convergent beam electron diffraction patterns. *Ultramicroscopy* **188**, 59–69.

Zhang, C., Feng, J., DaCosta, L. R. & Voyles, Paul. M. (2020). Atomic resolution convergent beam electron diffraction analysis using convolutional neural networks. *Ultramicroscopy* **210**, 112921.
Schnitzer 28